\documentstyle[12pt]{article}
\def\abstracts#1{{
\centering{\begin{minipage}{12.2truecm}\footnotesize
\baselineskip=12pt\noindent
\centerline{\footnotesize ABSTRACT}\vspace*{0.3cm}
\parindent=0pt #1
\end{minipage}}\par}}

\renewenvironment{thebibliography}[1]
{\begin{list}{\arabic{enumi}.}
{\usecounter{enumi}\setlength{\parsep}{0pt}
\setlength{\leftmargin 1.25cm}{\rightmargin 0pt}
 \setlength{\itemsep}{0pt} \settowidth
{\labelwidth}{#1.}\sloppy}}{\end{list}}
\newcounter{itemlistc}
\newcounter{romanlistc}
\newcounter{alphlistc}
\newcounter{arabiclistc}

 1
 1
 1

\newcommand{\ba}{\begin{array}}
\newcommand{\ea}{\end{array}}
\newcommand{\bd}{\begin{displaymath}}
\newcommand{\ed}{\end{displaymath}}
\newcommand{\be}{\begin{equation}}
\newcommand{\ee}{\end{equation}}
\newcommand{\bea}{\begin{eqnarray}}
\newcommand{\eea}{\end{eqnarray}}

\def\G{\Gamma}

\def\p{\pi}

\begin{document}
\vspace*{-1.8cm}
\begin{flushright}
{\normalsize IC/95/214}\\
{\normalsize MRI/PHY/18-95}
\end{flushright}
\centerline{\normalsize\bf SOME SIGNALS FOR A LIGHT NEUTRALINO  }
\vspace*{0.6cm}
\centerline{\normalsize
RATHIN ADHIKARI $^{1}$$^{*}$ }
\baselineskip=13pt
\centerline{\footnotesize\it International Centre for Theoretical
Physics}
\centerline{\footnotesize\it  P.O. Box 586, 34100, Trieste, Italy}
\centerline{\normalsize   and  }
\centerline{\normalsize
 BISWARUP MUKHOPADHYAYA }
\baselineskip=13pt
\centerline{\footnotesize\it Mehta Research Institute of Mathematics
and Mathematical Physics}
\centerline{\footnotesize\it  Allahabad - 211 002 , India}

\centerline{\footnotesize  E-mail: biswarup@mri.ernet.in}
\vspace*{0.9cm}
\baselineskip=12pt
\abstracts{
If a light gaugino sector exists in the supersymmetric standard model
then the mass of lightest neutralino may be
of the order of 1 GeV or less. As a consequence of neutral flavor violation
in supersymmetric theories $B_s$-meson may decay into a pair of lightest
neutralinos in such a case.
It is found that the parameter space for such light neutralinos can be
appreciably constrained by looking for such decays. We also show
how a rare B-decays ($B \longrightarrow K(K^{*}) + invisible$ channels)
can help us in probing a light neutralino in B-factories in a reasonably model-
independent manner. Finally, we observe that that the decay of a tau-lepton
into
a muon and a pair of light neutralinos can cause a violation of weak
universality which is larger in magnitude than that from any source known
so far.}

\vspace*{6mm}
\normalsize\baselineskip=15pt
\setcounter{footnote}{0}
\renewcommand{\thefootnote}{\alph{footnote}}

Although
the lower bound on the gluino mass in the  minimal supersymmetric (SUSY) model,
as obtained from hadronic collision experiments, is about 150 GeV \cite {cdf},
the stringency of the event selection criteria there allows a window~\cite
{pdg,lg1,lg2} in the range of 2.5 - 5 GeV, which cannot be unambiguously closed
even from low-energy phenomena.
Such a light gluino also relaxes the squark mass limits \cite {lg1}.
There are some theoretical motivations also for a light gluino from the
viewpoint of improved consistency in the running of the strong coupling
constant $\alpha_s$\cite {clav}. Naturally, such a situation also calls for
a small value for the mass of the lightest SUSY particle (LSP) which is the
lightest neutralino in most theories.
Furthermore, in this light gluino scenario such
a lightest neutralino is predominantly a photino in a  SUSY model embedded
in a Grand Unified Theory (GUT) ~\cite{GUT}. In such a case the
range
in the parameter space that is allowed by LEP experiments and is simultaneously
compatible with a light gluino corresponds to lightest neutralino mass
$\approx$ 0.5 to 1.5 GeV, $\mu
\approx -50 \;$ to $ - 100 \; $GeV
and $\tan \beta \approx 1.0-1.8$, $\mu$ and $\tan \beta$ being respectively
the Higgsino mass parameter and the ratio of the scalar vacuum
expectation \vadjust{
\vskip 2 mm
{\footnotesize  $^{1}$ Address after November, '95 :
Physical Research Laboratory, Ahmedabad,India. \\E-mail:
rathin@prl.ernet.in}

{\footnotesize $^{*}$  Presented at the International Symposium and
Workshops on
Particle Theory and Phenomenology, IITAP, Iowa State University, USA, May
17-26, 1995. \hfill} \pagebreak}

\noindent
values. Recently  a lightest stable neutralino in this mass range has been
claimed to be
consistent with astrophysical constraints in a special type
of SUSY model \cite {Kolb}.

Here we suggest some methods for exploring the parameter space of a scenario
containing a light neutralino. This discussion is model independent, except
that, to keep the
calculations simple and transparent, we have assumed the LSP
to be a photino following the guidelines of a GUT-based theory.

First we consider the two body decay of $B_s$
meson, namely, $B_{s}\longrightarrow  \chi^0_1 \chi^0_1$
where $\chi^0_1$ is the LSP \cite {phd}. Such an invisible decay of the
$B_s$ has no backgrounds in the standard model. At
the quark level, the the above decay  process
corresponds to $b\longrightarrow s \chi^0_1 \chi^0_1$. Interestingly, such
a flavor-changing neutral current (FCNC) process can be allowed at the
tree-level ~\cite{FCNC} in SUSY, due to a mismatch between the quark and squark
mass matrices in the left sector. The interaction involving
$b \longrightarrow s$ in this fashion is controlled by a term  $\G_{23}$,
$\G_{jk}$ being the (jk)-th element of the unitary matrix that
diagonalises $M_{\tilde{d}}^2$ where
\hspace{-2mm}
\bea
M_{L_{\tilde{d}}}^2 =
\left( m_L^2 \, {\bf{1}} + m_{\hat{d}}^2+c_0 \, K \, m_{\hat{u}}^2 \,
K^{\dag}
\right)
\eea

\noindent
The last term in $M_{\tilde{d}}^2$ is crucial here; it arises from
evolution of the squark mass parameter which receives corrections from
couplings of the charged Higgsinos. The value of $\G_{23}$
depends on $m_t$ and $c_0$. In view of the recent results from the
Fermilab Tevatron, we have chosen $m_{t} \; = \; 170 \;$ GeV here. The value
of $c_0$ is model dependent; however, as recent estimates indicate, a
value around 0.01 or slightly above is likely even  from a
\vskip 3.5 in
{\footnotesize Fig. 1. The branching ratio for invisible
$B_s$-decay (in units of
$c^2 \; f^{2}_{B_s}$) plotted against the LSP mass for
$m_{\tilde{q}}\;=\;80$ GeV }
\vskip 0.2 in

\noindent
rather conservative
point of view ~\cite{c0}. Here we write
$({\Delta m^{2}_{\tilde{q}} \over m^{2}_{\tilde{q}}})\G_{23} = c K_{23}$,
where c is treated as a phenomenological input, K is the Kobayashi-Maskawa
matrix and ${\Delta m^{2}_{\tilde{q}}}$ is the squark mass square
splitting. From rare decays
such as $b \rightarrow s \; \gamma $ ~\cite{gau}, a value of
$\mid c_0 \mid
 \approx 0.05$ is allowed for $m_t \approx 175 $ GeV and $m_{\tilde q}
\approx 60 $ GeV. For higher  $m_{\tilde q}$ this constraint
gets more
relaxed. In such cases $\Gamma_{23}$  is of the same order of magnitude
as $K_{23}$ . Thus for about 1\% splitting in squark masses, $c \approx 0.01$
 is easily possible.

The two-body decay-width shown in fig. (1) is
given by \hspace{-2mm}
\bea
\G \; = {g^4 \; \sin^4 \theta_w \; {|  K_{23}  | }^2 \; \left(  c^2
f^2_{B_s}  \right)  \over 216 \; \p \; {m_{\tilde q}}^4 } \; m^2 \;
{\left( \; {m_B}^2 \; - 4\; m^2 \right)}^{1/2}
\eea

\noindent
where $f_{B_s}$ is the $B_s$-decay constant,
and $m$ and $m_{\tilde{q}}$ are respectively the mass of LSP and the
average of the b-and s (left) squark
masses. In the light gluino scenario, $m_{\tilde{q}}$ = 80 GeV is within the
allowed region of the
parameter space. The branching ratio corresponding to
other values of $m_{\tilde q}$ can be obtained from the same graph using
eqn. (2) and with appropriate scaling.

In the graph, mass of LSP in the range 0.5 to 1.5 GeV
corresponds to a branching ratio of
$(10^{-3} - 10^{-2}) \; c^2 \; f^{2}_{B_s} \; {\mbox GeV}^{-2}$.
The value of the
parameter $f_{B_s}$, although not completely known yet, can
be expected to lie in the range of 0.3 GeV ~\cite{fBs}. Depending on this,
a branching
ratio of O$(10^{-4} - 10^{-3}) c^2$ can be expected for the invisible
channel.
If an accumulation  of $10^8$ $B
\overline{B}$-pairs takes place in a B-factory,
then the observation (or absence) of such decays could be employed to set
limits in the $m-c$ parameter space from the viewpoint of light LSP's.
This should be an
independent laboratory constraint, in addition to those obtained from, say,
decays  of light charginos which often occur in the light LSP
scenario. Moreover, if one
wants to ignore gaugino mass relations from GUT's and restrict
light LSP's from a purely phenomenological point of view, then it is possible
to put limits in the range of 1-2 GeV as well, the branching ratio
being even higher in that range.

Experimental observability of this invisible decay needs the efficiency
of reconstruction of one $B_s$ in the pair which is at present $O(10^{-3})$
 ~\cite{cleo}.  However, this efficiency can be increased to $O(10^{-2})$
by extending the
search techniques to decays like $B_{s} \longrightarrow D^{* \pm}_{s} X$
\cite{bort, stone}, taking into account both $\pi^{\pm}$ and $\rho^{\pm}$
as products,
and also using semileptonic tags.

We next consider the flavour-changing neutral current (FCNC)
three-body decays \\ $B\longrightarrow K(K^{*}) \chi^0_1 \chi^0_1$ \cite
{pll}. The energy spectrum of the $K(K^*)$ in this decay (which has the
same final state as that with $K (K^*)$ and neutrinos) shows an interesting
distortion  depending on the LSP mass.
At the quark level this decay has the same matrix element
as the earlier two-body decay process.
However, we need various form factors to express hadronic matrix elements
for the quark current.
Our results are based upon numerical values of the various form-factors
(and pole fits for their momentum-transfer dependence) obtained from the
relativistic quark model of reference \cite{wsb}. These form-factors have been
computed in the literature using other models, too \cite{bmod};
We find that the uncertainties in the values of the form-factors
do not destroy the general features of our results.

Also, the results to be shown below are susceptible to QCD corrections.
Though
such corrections moderately alter the decay rates \cite{qcd}, the key
featurs are not expected to be lost. This is because at the lowest
order electroweak level, the SUSY and standard model effective interactions
have the same operator structure, and our results depend on their relative
magnitudes.

To compute the energy distribution, one has to add
the differential decay rates for the
SUSY process with that for
$\Sigma B\longrightarrow K(K^{*}) \nu_i \overline{\nu_i}$ which occurs via
triangle as well as box diagrams \cite{soni}. The net observed variation of
$d\Gamma/dE_{K(K^{*})}$
with the K(K$^*$)-energy is a result of superposition of the two
types of final states, leading to a distribution with a kink.
The position of the kink and
the distortion to the spectrum relative to the purely SM case depends on
the mass of the LSP.\vadjust{
\vskip 3.5 in
 {\footnotesize Fig.2. The differential decay rates for $B
{\longrightarrow} K + nothing$
for $m_{\tilde{q}}\; = 100 GeV, \; c = 0.1$. The solid, dotted and short-dashed
curves correspond to three LSP masses expressed in GeV. The long-dashed
curve below is for the purely standard model case with three massless
neutrinos}
\vskip 0.3 in
}

The numerical results are shown in figures 2 for $K$ final states only.
We have drawn the graphs for $m_{\tilde{q}} = 100 GeV$ which is
easily allowed in this scenario and $c$, is treated here as a free
input parameter.  This enables us
to extend this study, if necessary, even beyond the minimal SUSY model.
Evidently, one can notice distortions to
the spectrum over a considerable region of the parameter space. The effect
becomes less and less obvious with increasing squark mass, and is
barely perceptible for $B\longrightarrow K \chi^0_1 \chi^0_1$ with
$c \approx .5$, $m_{\tilde{q}} = 500 GeV$ or $c \approx 0.05$,
$m_{\tilde{q}} = 100 GeV$. Also,
the response to a variation in the mass of the LSP
in the region  $0.5-1.5 GeV$ is manifest. A few hundred events in a B-factory
should suffice to explore this kind of a distortion .

It is to be noted that while the differential decay rate for
$\Sigma(B\longrightarrow K \nu_{i} \overline{\nu_{i}})$ increasess
monotonically  with $E_K$,
it dips after an initial rise in the case of
$\Sigma(B\longrightarrow K^{*} \nu_{i} \overline{\nu_{i}})$  and the
kinky characteristics of the distribution pattern is not so prominent for
the $K^*$ final states.
If $10^{7-8}$ $B \overline{B}$-pairs are produced in a B-factory
per year, then the above types of decays in B-factory experiments
are going to
help one in constraining the light sparticle scenario to a large extent.

  As a digression, it may be mentioned that the same spectral distortion
as the one described above occurs in the minimal SUSY
standard model in a general scenario also. The process in question is the
decay $H \longrightarrow Z + invisible $ where one has to add the
contributions from $Z $ and pairs of lightest neutralinos as well as
$Z$ and neutrinos (three massless species) as final decay products. Here
also we see the high sensitivity of the neutralino mass in the kinky
characteristics of the differential decay width distributions against
$Z$-energy \cite {biswa} which would otherwise have had a uniform rise
due to the neutrino contributions alone. This feature is visible for the
LSP mass in the range 150 - 200 GeV, for the decay of a Higgs
having mass 500 GeV or so.

  Lastly we like to mention that for the lightest neutralino in the range
of a few hundred MeV, the decay
$\tau \longrightarrow \mu \chi^0_1 \chi^0_1 $ is also allowed and this
leads to the violation of tau-universality \cite{tau}. Here, again the
flavour
violation is  controlled by an effect of non-diagonal corrections to
the slepton mass matrix, and is favored in models with massive (Majorana?)
neutrinos. It is found that
this violation can be greater than both non-universal electroweak
radiative corrections and supersymmetric one-loop corrections over a
considerable region of SUSY parameter space allowed by experiments so far.
Thus in addition to B-factories, tau factories may also be quite helpful in
either constraining the parameter space for lightest neutralino in the low
mass region or in finding it.

\vspace*{6mm}
\noindent
{\bf Acknowledgements}

             R. A. likes to thank the High Energy Physics section,
International Centre for Theoretical Physics, Trieste, Italy for its
hospitality during writing this and IITAP, Iowa State University for
invitation to present these works.


\end{document}